\documentclass[submission,copyright,creativecommons]{eptcs}
\providecommand{\event}{FOCLASA 2012} 
\usepackage{breakurl}             
\usepackage{graphicx}
\usepackage{tabularx}

\usepackage{subfig}
\usepackage{times}
\usepackage{verbatim}
\usepackage{comment}
\title{A Case Study on Formal Verification of Self-Adaptive Behaviors in a Decentralized System}
\author{M. Usman Iftikhar \qquad\qquad Danny Weyns
\institute{School of Computer Science, Physics and Mathematics\\
Linnaeus University,  V\"axj\"o, Sweden}
\email{usman.iftikhar@lnu.se, danny.weyns@lnu.se}
}
\def\titlerunning{A Case Study on Formal Verification of Self-Adaptive Behaviors in a Decentralized System}
\def\authorrunning{M. Usman Iftikhar \& Danny Weyns}

\begin{document}
\newcommand{\comments}[1]{} 
\maketitle

\begin{abstract}

Self-adaptation is a promising approach to manage the complexity of modern software systems. A self-adaptive system is able to adapt autonomously to internal dynamics and changing conditions in the environment to achieve particular quality goals. Our particular interest is in decentralized self-adaptive systems, in which central control of adaptation is not an option. One important challenge in self-adaptive systems, in particular those with decentralized control of adaptation, is to provide guarantees about the intended runtime qualities. In this paper, we present a case study in which we use model checking to verify behavioral properties of a decentralized self-adaptive system. Concretely, we contribute with a formalized architecture model of a decentralized traffic monitoring system and prove a number of self-adaptation properties for flexibility and robustness. To model the main processes in the system we use timed automata, and for the specification of the required properties we use timed computation tree logic. We use the Uppaal tool to specify the system and verify the flexibility and robustness properties.

\end{abstract}

\section{Introduction}

Our society extensively relies on the qualities of software systems, e.g., the reliability of software for media, the performance of software for manufacturing, and the openness of software for enterprise collaborations. However, ensuring the required qualities of software that has to operate in dynamic environments poses severe engineering challenges.  
Self-adaptation is generally considered as a promising approach to manage the complexity of modern software systems~\cite{Kep:Che,kramer2007sms,cheng:roadmap,Lemos2012}. Self-adaptation enables a system to adapt itself autonomously to internal dynamics and changing conditions in the environment to achieve particular quality goals. It is widely recognized that software architecture provides the right level of abstraction and generality to deal with the challenges of self-adaptation~\cite{Oreizy1998,Rainbow,kramer2007sms}. 
In particular, the use of an architecture-based approach 
can provide an appropriate level of abstraction to describe dynamic change in a system, such as the use of components, bindings and composition, rather than at the algorithmic level.
Our particular interest is in decentralized self-adaptive systems, in which central control of adaptation is not an option. 
Examples are large-scale traffic systems, integrated supply chains, and federated cloud infrastructures.

One important challenge in self-adaptive systems, in particular those with decentralized control of adaptation, is to provide guarantees about the required runtime quality properties.  In previous research, we have defined formally founded design models for decentralized self-adaptive systems that cover \textit{structural} aspects of self-adaptation~\cite{Weyns2011}. These models support engineers with reasoning about structural properties, such as types and interface relations of different parts of the decentralized system. However, in order to provide guarantees about qualities, we need to complement this work with an approach to validate \textit{behavioral} properties of decentralized self-adaptive systems. The need for research on formal verification of behavioral properties of self-adaptive systems is broadly recognized by the community~\cite{Magee2006,Zhang2006,Vassev2009,Lemos2012}. 

This paper reports a first step of our research goal to develop an integrated approach to validate behavioral properties of decentralized self-adaptive systems to guarantee the required qualities~\cite{WODA2012}. This approach integrates three activities: (1) model checking of the behavior of a self-adaptive system during design, (2) model-based testing of the concrete implementation during development, and (3) runtime diagnosis after system deployment. 
The key underlying idea of the approach is to enhance validation of qualities by transfering formalization results over different phases of the software life cycle, e.g., model based testing starts with a verified model and a set of required properties and then intends to show that the implementation of the system behaves compliant with this model. The focus of this paper is on the first activity. 
Concretely, we present a case study of  a decentralized traffic monitoring system and use model checking to guarantee a number of self-adaptation properties for flexibility and robustness. With flexibility we refer to the ability of the system to adapt dynamically with changing conditions in the environment, and robustness is the ability of the system to cope autonomously with errors during execution.  We model the main system processes with timed automata and specify the required properties using timed computation tree logic (TCTL). We use the Uppaal tool that offers an integrated environment for modeling, simulation and verification, based on automata and a subset of TCTL.

The remainder of this paper is structured as follows. In Section~\ref{sec:case}, we introduce the traffic monitoring system and explain a number of adaptation scenarios. In Section~\ref{sec:approach}, we give a brief background on formal modeling with Uppaal. Section~\ref{sec:design} presents the design model of the traffic monitoring system, and Section~\ref{sec:verification} explains how we verified key properties and discusses potential uses of the study results both as input for model based testing and as a starting point for the definition of a reusable behavior model for self-adaptive systems. We discuss related work in Section~\ref{sec:relatedwork}, and conclude with a summary and challenges ahead in Section~\ref{sec:conclusions}.

\section{Traffic Monitoring System}\label{sec:case}

Intelligent transportation systems (ITS) is a worldwide initiative to exploit information and communication technology to improve traffic.\footnote{http://ec.europa.eu/transport/its/, http://www.its.dot.gov/} One of the challenges in this area is effective monitoring of traffic.  In \cite{MACODO-arch}, we have introduced a monitoring system that provides information about traffic jams. This information can be used to reduce traffic congestion by different types of clients, such as traffic light controllers, driver assistance systems, etc.  The main challenges of the system are: (1) inform clients of dynamic changing traffic jams, (2) realize this functionality in a decentralized way, avoiding the bottleneck of a centralized control center, (3)  make the system robust to camera failures. Whereas the focus in \cite{MACODO-arch} was on the structural aspects, here we focus on the behavioral aspects of the system' architecture. 

\begin{figure}[th!]
	\centering
    \includegraphics[width=0.72\textwidth]{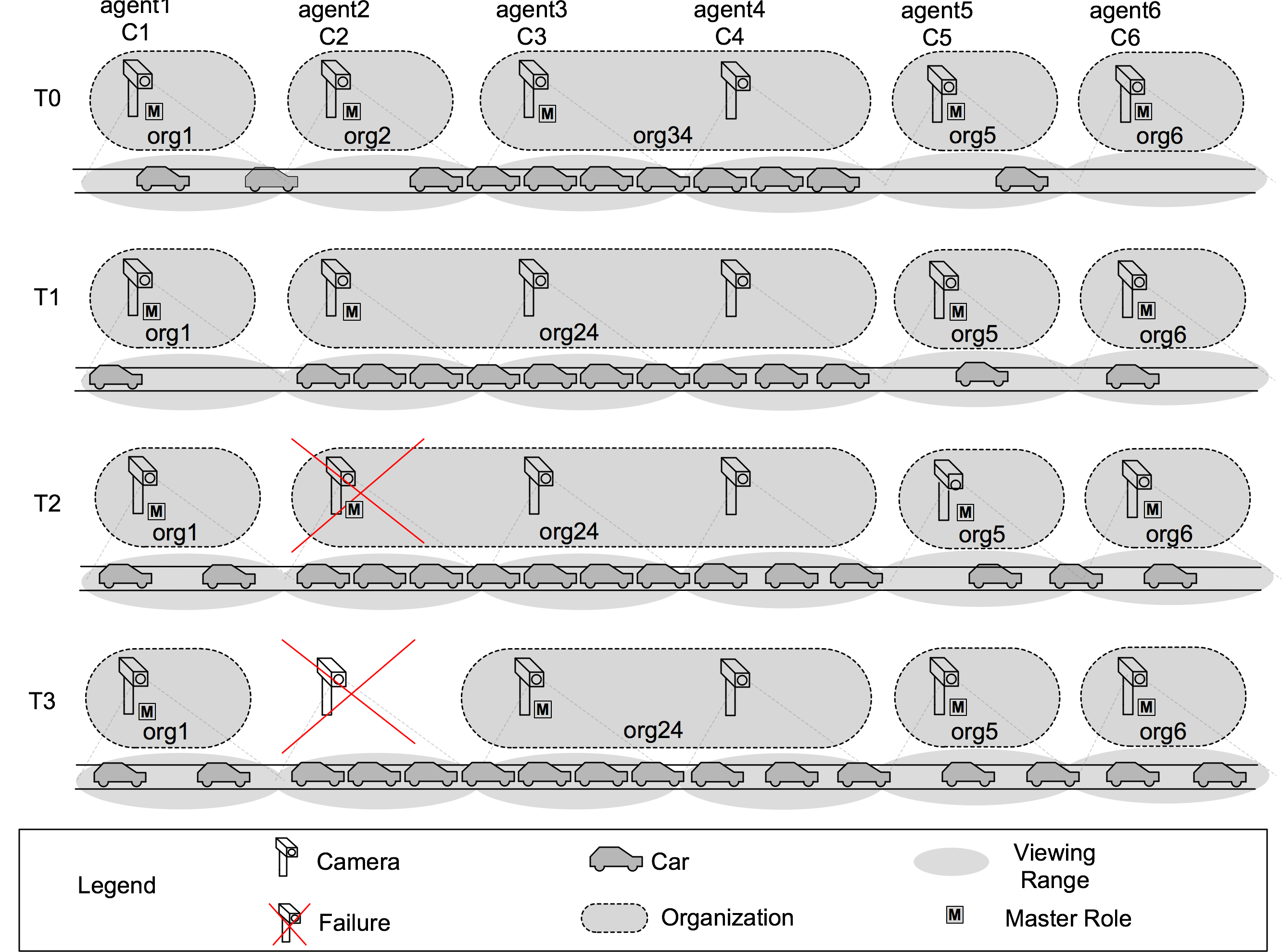}
     \caption{Self-healing scenario}   \label{fig:scenario}
         \vspace{-8pt}
\end{figure}

The system consists of a set of intelligent cameras, which are distributed along the road.~An example of a highway is shown in Fig.~\ref{fig:scenario}. 
Each camera has a limited viewing range and cameras are placed to get an optimal coverage of the highway with a minimum overlap.  To realize a decentralized solution, cameras collaborate in organizations: if a traffic jam spans the viewing range of multiple cameras, they form an organization that provides information to clients that have an interest in traffic jams. 

Fig.~\ref{fig:scenario} shows two scenarios that require adaption. The first scenario concerns the dynamic adaptation of an organization from T0 to T1, where camera 2 joins the organization of cameras 3 and 4 after it monitors a traffic jam. The second scenario concerns robustness to a silent node failure, i.e.,~a failure in which a failing camera becomes unresponsive without sending any incorrect data. This scenario is shown from T2 to T3, where camera 2 fails. Since there are dependencies between the software running on different cameras (details below), such failures may bring the system in an inconsistent state and disrupt its services. Therefore, the system should be able to restore its services after a failure, although in degraded mode since the traffic state is no longer monitored in the viewing range of the failed camera.

\subsection{Dynamic Agent Organizations for Flexibility}\label{sec:arch}

\begin{figure*}
\centering
  \subfloat[Primary software components deployed on each camera]{
	\label{fig:toplevel}
	\includegraphics[width=0.38\textwidth]{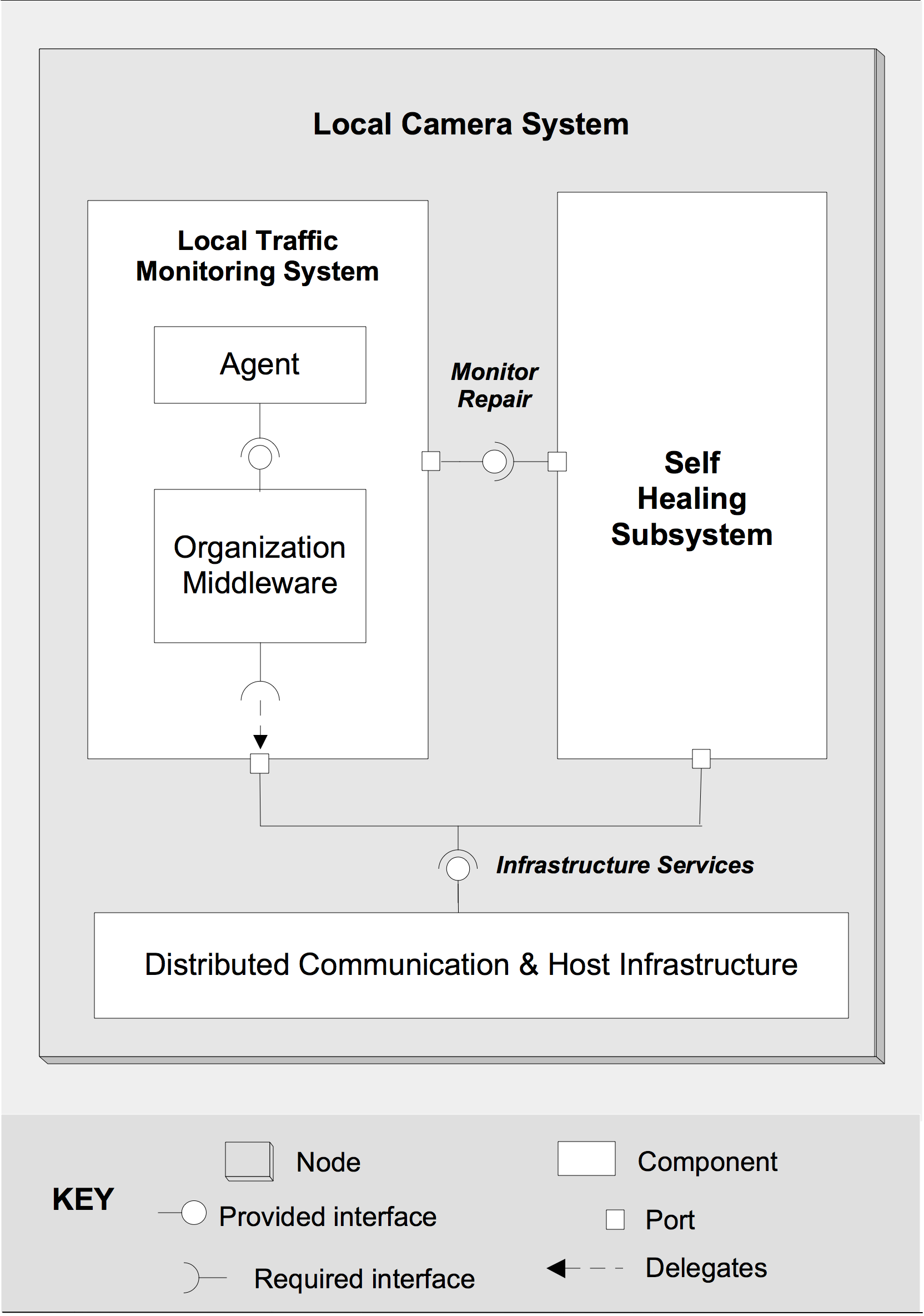}}
 \hspace{20pt}
 \subfloat[Design model of camera and environment (communication channels are omitted)]{
	\label{fig:toplevelmodel}
	\includegraphics[width=0.4\textwidth]{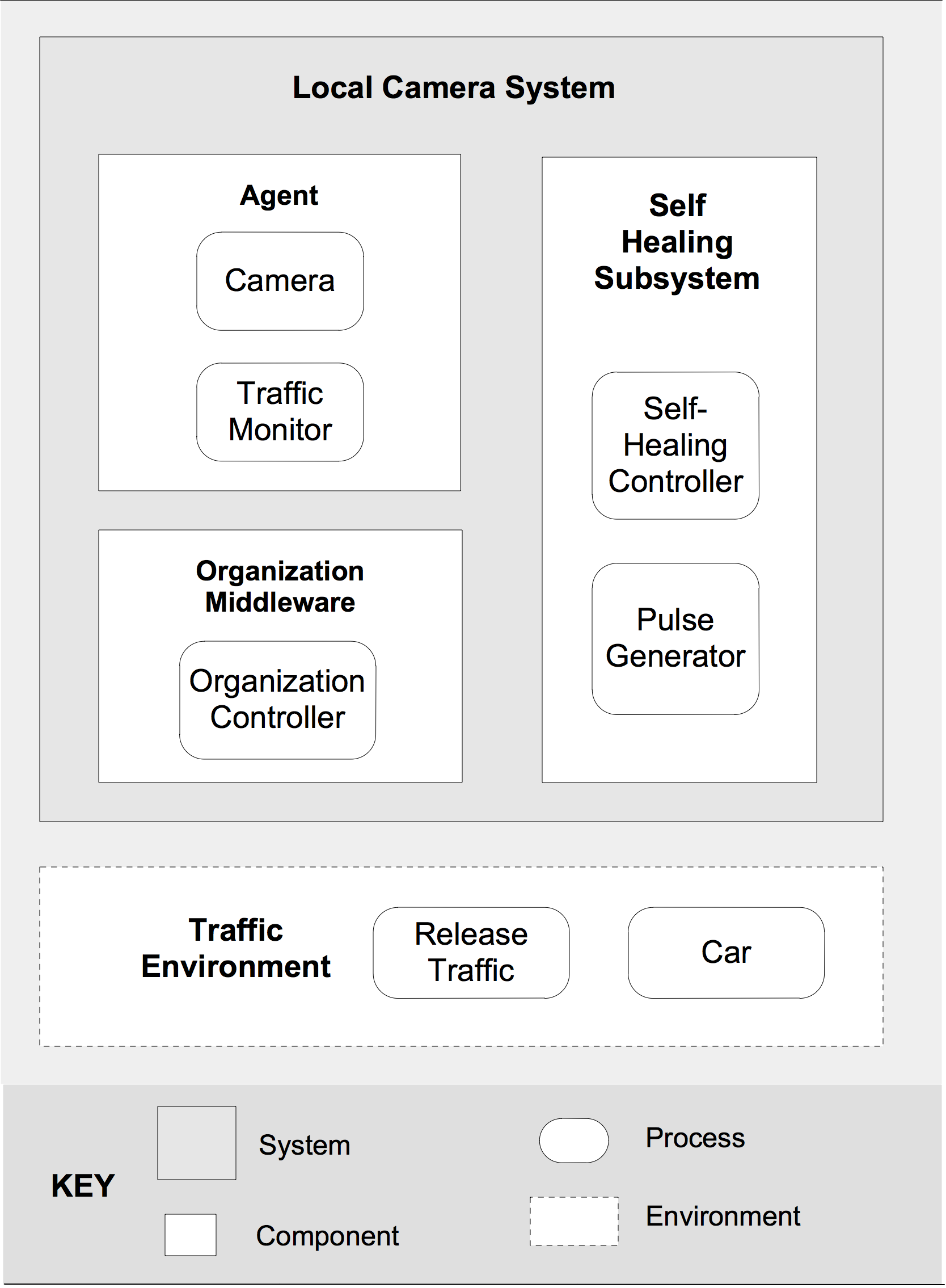}}
 \caption{Local Camera System}
 \label{fig:framework}\vspace{-10pt}
\end{figure*}

Figure~\ref{fig:toplevel} shows the primary components of the software deployed on each camera, i.e.~the \emph{local camera system}. The \emph{local traffic monitoring system} provides the functionality to detect traffic jams and inform clients. The local traffic monitoring system is conceived as an agent-based system consisting of two components. The \emph{agent} is responsible for monitoring the traffic and collaborating with other agents to report a possible traffic jam to clients. The \emph{organization middleware} offers life cycle management services to set up and maintain organizations. We employ dynamic organizations of agents to support flexibility in the system, that is, agent organizations dynamically adapt themselves with changing traffic conditions. To access the hardware and communication facilities on the camera, the local traffic monitoring system can rely on the services provided by the \emph{distributed communication and host infrastructure}. 

In normal traffic conditions, each agent belongs to a single member \emph{organization}. However, when a traffic jam is detected that spans the viewing range of multiple neighboring cameras, organizations on these cameras will merge into one organization. To simplify the management of organizations and interactions with clients, the organizations have a master/slave structure. The master is responsible for managing the dynamics of that organization (merging and splitting) by synchronizing with its slaves and neighboring organizations and reporting traffic jams to clients. Therefore, the master uses the context information provided by its slaves about their local monitored traffic conditions. At T0, the example in Fig.~\ref{fig:scenario} shows four single member organizations, \emph{org1} with \emph{agent1}, \emph{org2} with \emph{agent2}, and similar for \emph{org5}, and \emph{org6}. Furthermore, there is one merged organization, \emph{org34} with \emph{agent3} as master and \emph{agent4} as slave. At T1, the traffic jam spans the viewing range of cameras 2, 3 and 4. As a result, organizations \emph{org2} and \emph{org34} have merged to form \emph{org24} with \emph{agent2} as master. When the traffic jam resolves, the organization is split dynamically. 

\subsection{Self-Healing Subsystem for Robustness}

To recover from camera failures, a \emph{self-healing subsystem} is added to the local traffic monitoring system, as shown in Fig.~\ref{fig:toplevel}. The self-healing subsystem maintains a model of the current dependencies of the components of the local traffic monitoring system with other active cameras. 
Each working camera is in one of three distinct roles: master of a single member organization, master of an organization with slaves, or slave in an organization. As these roles come with certain responsibilities, each camera is dependent on a particular set of remote cameras in order to function properly: 
%\begin{list}{\labelitemi}{\leftmargin=1.25em}
%	\item[1.] 
(1) a master of a single agent organization is dependent on its neighboring nodes; 
%	\item[2.] 
(2) a master with slaves is dependent on its slaves and its neighboring nodes;
%	\item[3.] 
(3) a slave of an organization is dependent on its master and its neighboring nodes.
%\end{list}

To recover from camera failures, the subsystem contains repair actions for failure scenarios in different roles. Examples of actions are: halt the communication with the failed neighboring camera, elect a new master, and exchange the current monitored traffic state with another camera. To detect failures, the self-healing subsystem coordinates with self-healing subsystems on other cameras in the dependency model using a ping-echo mechanism. Cameras send periodically ping messages to dependent cameras and a failure is detected when a camera does not respond with an echo after a certain time.  
\cite{MACODO-arch}~provides a detailed description of the structural architecture of the traffic monitoring system. 

\section{Uppaal}\label{sec:approach}
Model-checking is verifying a given model w.r.t. a formally expressed requirement specification.
~Uppaal is a model-checking tool for verification of behavioral properties~\cite{Behrmann2006}. In Uppaal, a system is modeled as a network of timed automata, called processes. A timed automaton is a finite-state machine extended with clock variables. A clock variable evaluates to a real number, 
and clocks progress synchronously.  It is important to note that fulfilled constraints for the clock values only enable state transitions but do not force them to be taken. A process is an instance of a parameterized template. A template can have local declared variables, functions, and labeled locations. The syntax to declare functions is similar to that of the C language. State of the system is defined by locations of the automata, clocks, and variables values.

Uppaal uses a subset of TCTL for defining requirements, called the query language. The query language consists of state formulae and path formulae. State formulae describe individual states with regular expressions such as $x >= 0$. State formulae can also be used to test whether a process is in a given location, e.g., \emph{Proc.loc}, where \emph{Proc} is a process and \emph{loc} is a location.
Path formulae quantify over paths of the model and can be classified into \emph{reachability}, \emph{safety}, and \emph{liveness} properties: 
\begin{list}{\labelitemi}{\leftmargin=1.25em}
	\item[$\bullet$] \emph{Reachability} properties are used to check whether a given state formula $\phi$ can be satisfied by some reachable state. The syntax for writing this property is $E<>\,\phi$.
	\item[$\bullet$] \emph{Safety} properties are used to verify that ``something bad will never happen.'' There are two path formulae for checking safety properties. $A[]\ \phi$ expresses that a given state formula $\phi$ should be true in all reachable states, and $E[]\ \phi$ means that there should exist a path that is either infinite, or the last state has no outgoing transitions, called maximal path, such that $\phi$ is always true.
	\item[$\bullet$] \emph{Liveness} properties are used to verify that something eventually will hold, which is expressed as $A<>\ \phi$. The property ``whenever $\phi$ holds, eventually $\psi$ will happen'' is stronger and is expressed as $\phi \rightarrow \psi$.
\end{list}

Processes communicate with each other through \emph{channels}.  \emph{Binary channels} are declared as $chan\ x$. The sender $x!$ can synchronize with the receiver $x?$ through an edge. If there are multiple receivers $x?$ then a single receiver will be chosen non-deterministically. The sender $x!$ will be blocked if there is no receiver. \emph{Broadcast channels} are declared as $broadcast\ chan\ x$. The syntax for sender $x!$ and receiver $x?$ is the same as for binary channels. However, a broadcast channel sends a signal to all the receivers, and if there is no receiver, the sender will not be blocked. Uppaal also supports arrays of channels. The syntax to declare them is $chan\ x[N]$ or $broadcast\ chan\ x[N]$, and sending and receiving signals are specified as $x[id]!$ and $x[id]?$. Note that processes cannot pass data through signals. If a process wants to send data to another process then the sender has to put the data in a shared variable before sending a signal and the receiver will get the data from shared variable after receiving the signal.

Uppaal offers a graphical user interface (GUI) and model checking engine. The GUI consists of three parts: the editor, the simulator and the verifier. The editor is used to create the templates. The simulator is similar to a debugger, which can be used to manually run the system, showing running process, their current location and the values of the variables and clocks. The verifier is used to check properties of the model as described above.

\section{Model Design in Uppaal}\label{sec:design}

We now discuss the formal design of the traffic monitoring system in Uppaal. We already explained in section~\ref{sec:case} that each camera consists of 3 primary components: \emph{Agent}, \emph{Organization Middleware} and \emph{Self-healing Subsystem}. We have designed each of these components as a set of timed automata (templates) that represent abstract processes. Fig.~\ref{fig:toplevelmodel} (section~\ref{sec:arch}) shows how the processes map to the components of the system. To instantiate a particular system model, each template is instantiated to one or several concrete processes. We use channels to enable processes to communicate within a camera and between cameras. To that end, the id of the receiver camera is used.  We start by defining the different templates of the system. Then we explain how the templates are instantiated into a concrete system model. 

\subsection{Environment Processes}
The environment is modeled as two simple timed automata: \emph{Release Traffic} and \emph{Car}.  

\begin{figure*}
\centering
  \subfloat[Release traffic]{
	\label{fig:trafficSystem}
	\includegraphics[width=0.28\textwidth]{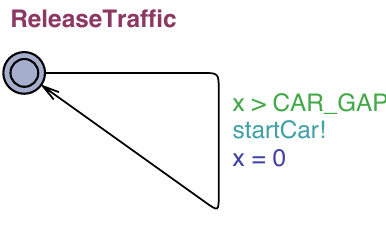}}
 \hspace{5pt}
 \subfloat[Car]{
	\label{fig:car}
	\includegraphics[width=0.62\textwidth]{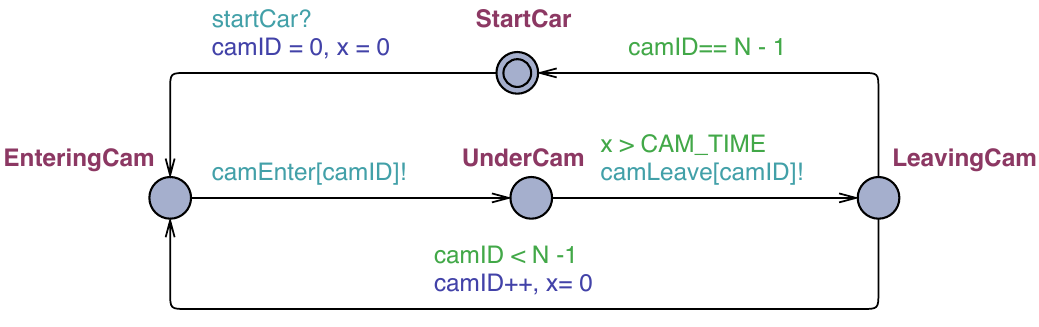}}
 \caption{Environment processes}
 \label{fig:traffic_car}\vspace{-5pt}
\end{figure*}

\begin{list}{\labelitemi}{\leftmargin=1.25em}
\item[$\bullet$] \emph{Release Traffic} is an abstract model of the traffic environment. Fig.~\ref{fig:trafficSystem} shows the template.  The purpose of traffic release is to feed the system with cars after some non-deterministic time $\mathit{CAR\_GAP}$. Variable $x$ is a local clock and whenever its value is greater than $\mathit{CAR\_GAP}$, 
the $\mathit{StartCar}$ signal is emitted. Only one instance of the release traffic process will be running all the time.
\end{list}
\begin{list}{\labelitemi}{\leftmargin=1.25em}
\item[$\bullet$] \emph{Car}  is the abstract model of a car in the environment. Fig.~\ref{fig:car} shows the template. $Car$ waits for the $\mathit{startCar}$ signal from the release traffic process.\footnote{$\mathit{startCar}$ is the initial location marked by two circles.} Once started, the car moves along the subsequent viewing ranges of the cameras. Whenever a car enters/leaves the viewing range of a particular camera it emits a signal. This allows the camera agents to monitor traffic congestion. As in  real traffic, the car template will have many running instances, each representing a car in the system.
\end{list}

\subsection{Agent Processes}
\noindent
An agent is modeled as two timed automata: \emph{Camera} and \emph{Traffic Monitor}.  
\begin{list}{\labelitemi}{\leftmargin=1.25em}
\item[$\bullet$] Each \emph{Camera} has four basic states. In normal operation, the camera can be master with no slaves, master of an organization with slaves, or it can be slave. Additionally, the camera can be in the failed state, representing the status of the camera after a silent node failure. Fig.~\ref{fig:camera} shows the template. There is an instance with a unique id for each camera. Cameras in the master status are responsible for communicating the traffic conditions to clients, but this functionality is not modeled here. 

\begin{figure*}
\centering
  \subfloat[Camera]{
	\label{fig:camera}
	\includegraphics[width=0.54\textwidth]{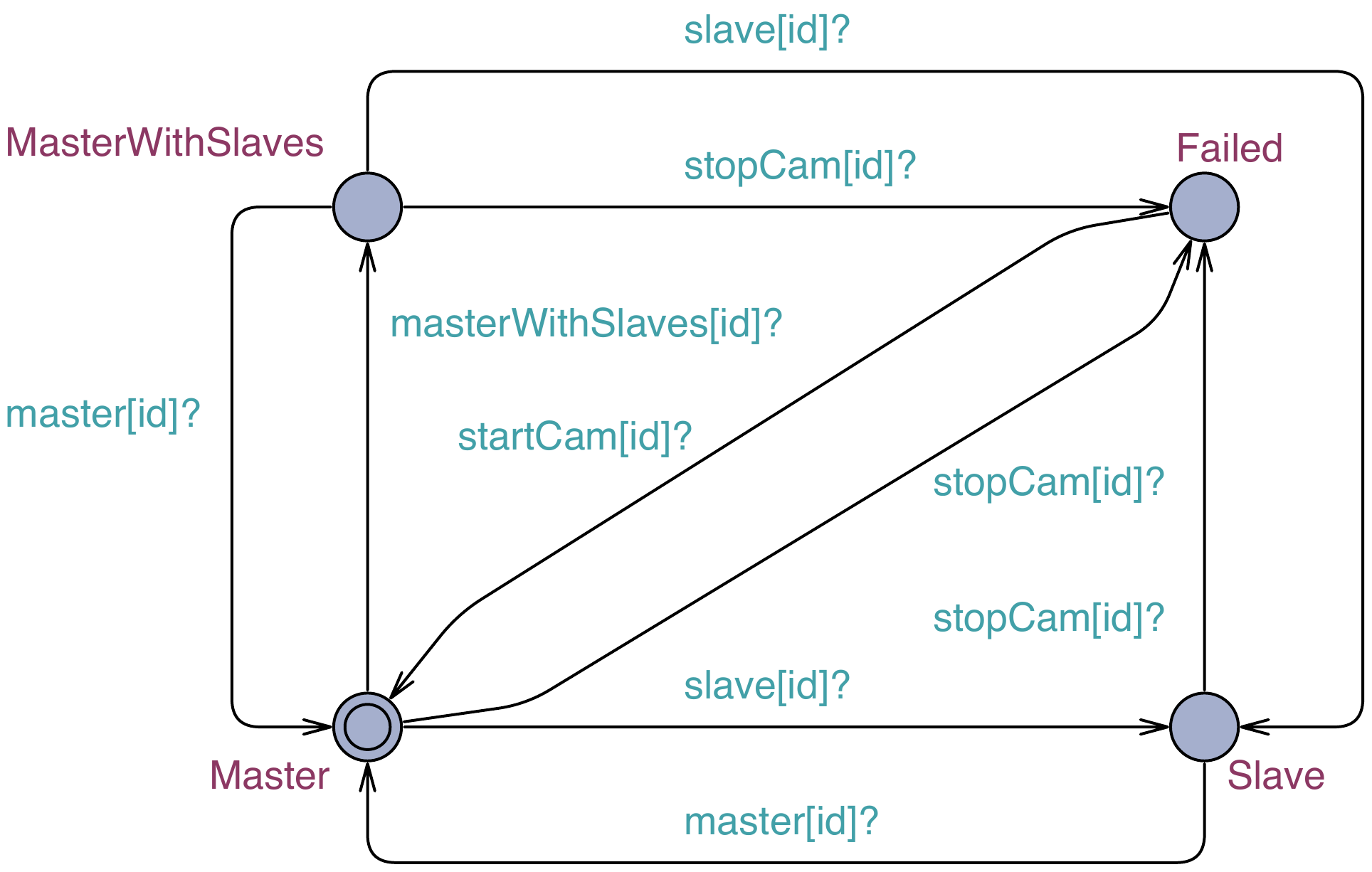}}
 \subfloat[Traffic monitor]{
	\label{fig:agent}
	\includegraphics[width=0.44\textwidth]{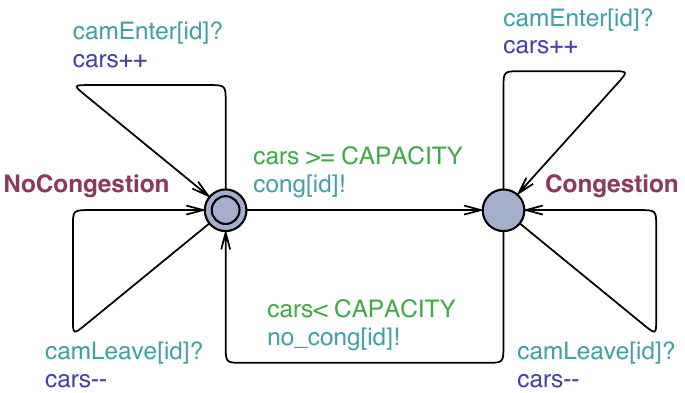}}
 \caption{Agent processes}
 \label{fig:camera_agent}\vspace{-5pt}
\end{figure*}

\item[$\bullet$] \emph{Traffic Monitor} keeps track of the actual traffic conditions based on the signals it receives from the cars and determines traffic congestion. It interacts with the local organizational manager to handle organization management. Fig.~\ref{fig:agent} shows the template. For each camera, one traffic monitoring process instance is running all the time.
Whenever a car enters into the viewing range of a camera, the traffic monitor detects the car via the $\mathit{camEnter}$ channel. Similarly when a car goes out of the range of a camera, the traffic monitor detects this through the $\mathit{camLeave}$ channel. The traffic monitors determines a traffic jam by comparing the total number of cars in its viewing range with the $\mathit{CAPACITY}$. Based on this, the monitor may interact with the organization controller to adapt the organizations.
\end{list}

\subsection{Organization Middleware}
The organization middleware is modeled as one timed automaton: \emph{Organization Controller}.  
\begin{list}{\labelitemi}{\leftmargin=1.25em}
\item[$\bullet$] \emph{Organization Controller} is responsible for managing organizations, based on the information it gets from the traffic monitor process of the agent. Fig.~\ref{fig:orgMiddleware} shows the template.\footnote{Committed states are marked with C. These states cannot delay, and the next transition must involve an outgoing edge of at least one of the committed locations.} An organization middleware process runs on each camera.
\begin{figure*}[th!]
    \centering	 \includegraphics[width=.95\textwidth, angle=90, totalheight=.92\textheight]
{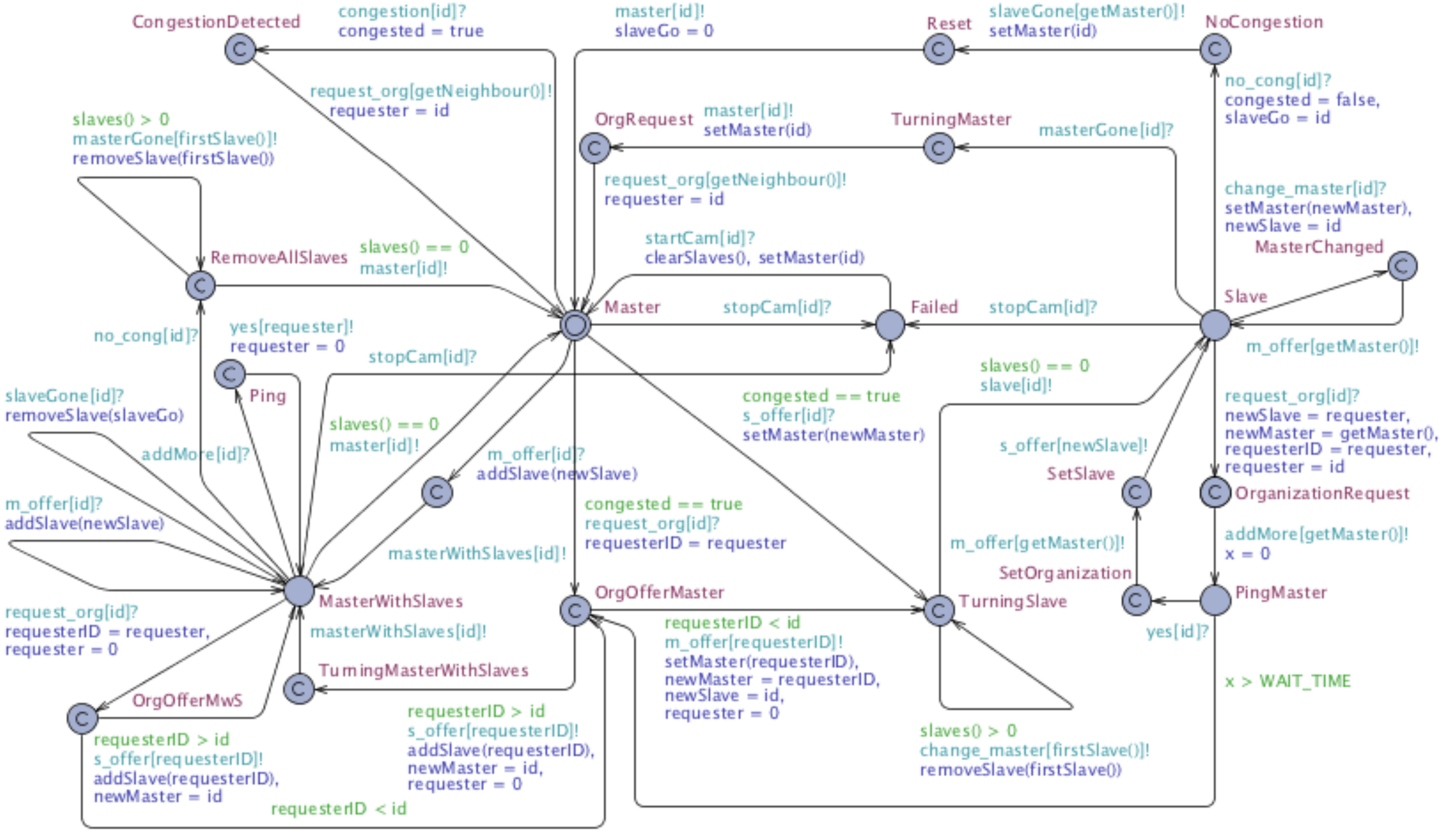}
 	\caption{Organization controller}\label{fig:orgMiddleware}
     \vspace{-8pt}
\end{figure*}
A camera starts as master of a single member organization. 
When a $\mathit{congestion}$ is detected the organization controller sends 
a $\mathit{request\_org}$ 
signal to the neighboring camera in the direction of the traffic flow. Depending on the traffic conditions of the neighboring camera and the current role of the camera, the organizations may be restructured as follows. If traffic is not jammed in the viewing range of the neighbor, organizations are not changed. If traffic is jammed and the neighbor is $\mathit{Slave}$ or $\mathit{MasterWithSlaves}$, the camera joins the organization as a slave. If both are masters of single member organizations ($\mathit{Master}$), the camera with the highest id becomes master with slaves of the joined organization (transition $\mathit{OrgOfferMaster}$ to $\mathit{TurningMasterWithSlaves}$ to $\mathit{MasterWithSlaves}$), while the other will become slave (transition $\mathit{OrgOfferMaster}$ to $\mathit{TurningSlave}$ to $\mathit{Slave}$). A master with slaves can add and remove slaves dynamically. When no slave remains, the master with slaves becomes master again. 
Whenever, the role of a camera changes, the organization controller informs the camera process to update its status 
via signals $\mathit{slave[id]}$,  $\mathit{master[id]}$, or  $\mathit{masterWithSlaves[id]}$ respectively. 
If the organization controller receives the $\mathit{stopCam}$ signal, it will go to $\mathit{Failure}$ state, which represents a silent node failure. The controller will not respond until it is recovered via the $\mathit{startCam}$ signal. 
\end{list}

\subsection{Self-Healing Subsystem}
\begin{figure*}
\centering
  \subfloat[Self-healing controller]{
	\label{fig:selfHealingSubsystem}
	\includegraphics[width=0.5\textwidth]{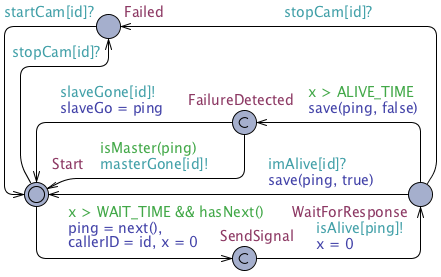}}
 \hspace{18pt}
 \subfloat[Pulse generator]{
	\label{fig:pulseGenerator}
	\includegraphics[width=0.34\textwidth]{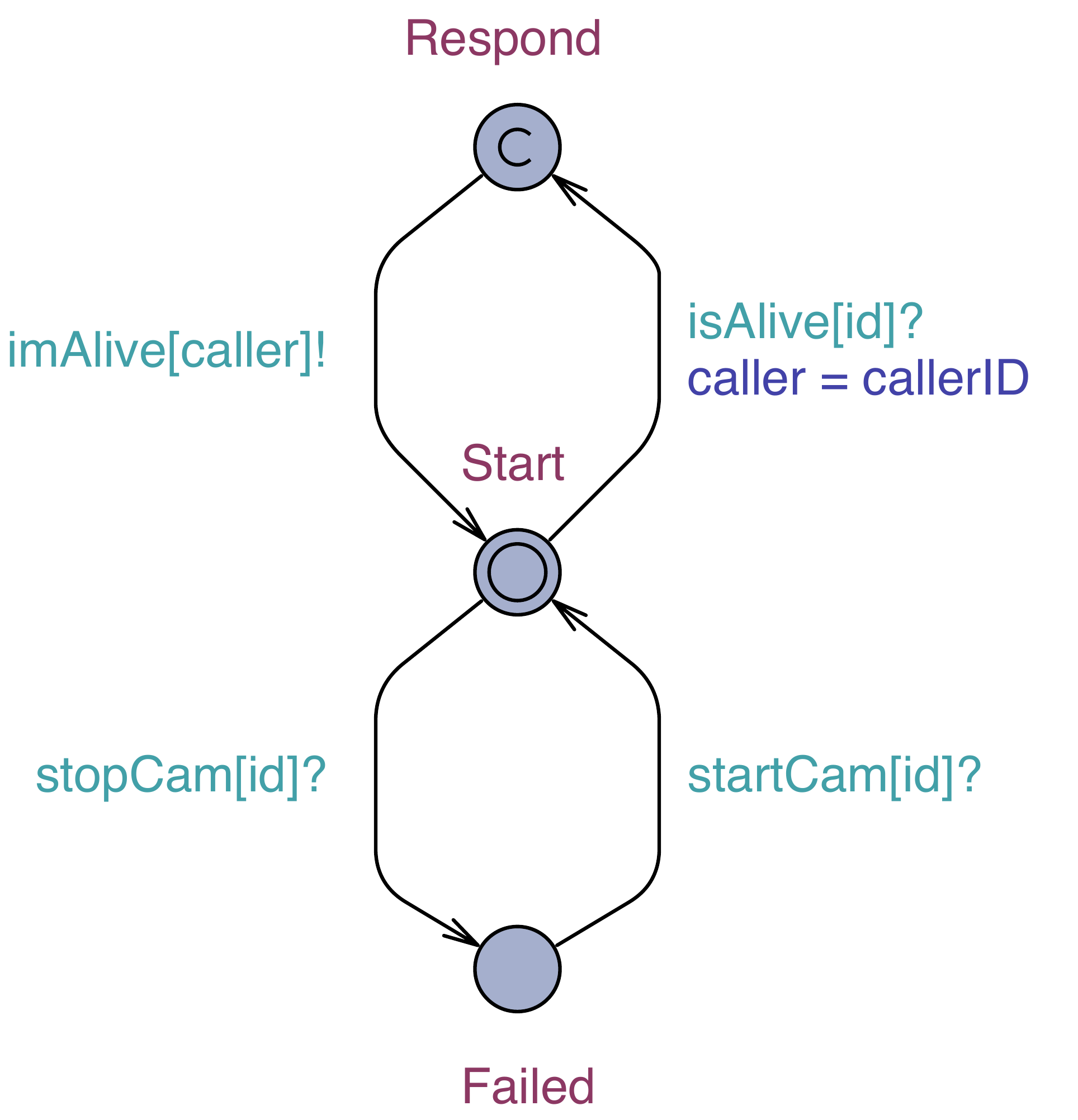}}
 \caption{Processes self-healing subsystem}
 \label{fig:self_healing}\vspace{-5pt}
\end{figure*}

The self-healing subsystem is modeled as two automata: \emph{Self-Healing Controller} and \emph{Pulse Generator}.  
\begin{list}{\labelitemi}{\leftmargin=1.25em}

\item[$\bullet$] \emph{Self-Healing Controller} is used to detect failures of other cameras based on a ping-echo mechanism. Fig.~\ref{fig:selfHealingSubsystem} shows the template. A self-healing controller process runs on each camera. The self-healing controller sends periodically $\mathit{isAlive[ping]}$ signals (based on $\mathit{WAIT\_TIME}$) to the self-healing controllers of the dependent cameras. If a camera does not respond in a certain time ($\mathit{ALIVE\_TIME}$) it adapts the organizational controller, either by removing a dependency in case a slave failed, or by restructuring the organization in case the master of the organization failed.

\item[$\bullet$] \emph{Pulse Generator} is responsible to respond to the ping signals sent by other cameras to check whether a particular camera is alive or not. Fig.~\ref{fig:pulseGenerator} shows the template.
\end{list}

\subsection{Definition Model Instance}\label{sec:instance}

A concrete system model is defined in Uppaal's system declarations section by listing the processes that have to be composed into a system: 

\small
\begin{verbatim}
   system Camera, TrafficMonitor, OrganizationController, 
          SelfHealingController, PulseGenerator, 
          ReleaseTraffic, Car;
\end{verbatim}
\normalsize
To define a concrete model, each template has to be instantiated to concrete processes. The process instances are defined in Uppaal's project declarations section: 

\small
\begin{verbatim}
   const int N = 6;         // # Camera
   typedef int[0, N-1] cam_id;
   ...
   // Global Constants
   const int CAM_TIME = 10;
   const int RECOVER_TIME = 500;
   ... 
   // Channels
   chan startCar;
   chan camEnter[N], camLeave[N], no_cong[N];
   ...
   chan reqTrafficJam, repTrafficJam[N];
   broadcast chan request_org[N], change_master[N], congestion[N];
   ...
\end{verbatim}
\normalsize
The declaration defines a setup with 6 camera systems.  As each local camera system has 5 processes (see Fig.~\ref{fig:framework}), we will have 30 processes running, plus the environment processes. Unique identifiers  are used to identify related processes to work together.  We have used an array of channels to enable processes to communicate within each camera and between cameras. If a process has to communicate with another process of the same camera it uses the id of the camera. If a process wants to communicate with a process of another camera it will use the id of the other camera. 

\section{Model Checking}\label{sec:verification}

Based on the formal design of the traffic monitoring system, we can now check properties of the traffic monitoring system by Uppaal's verifier. We have divided the properties in 3 groups, respectively system invariants, correctness of dynamic organization adaptations, and correctness of robustness to silent node failures.  We start by defining the properties. Then, we discuss the design and verification process. 

\subsection{System Invariants}

The following properties should hold: 

\begin{list}{\labelitemi}{\leftmargin=1.25em}
	\item[$\bullet$] All cameras cannot be slave at the same time (the system would not provide its function, that is, there are no masters that inform clients about traffic congestion):
\end{list}
\small
\begin{verbatim}
I1:   A[] not forall(i: cam_id) Camera(i).Slave
\end{verbatim}
\normalsize
\begin{list}{\labelitemi}{\leftmargin=1.25em}
	\item[$\bullet$] The system should not be in deadlock at any time. A deadlock may for example occur when the self-healing controllers are waiting on each other for responses to ping messages and none of them is able to send a response. Such situations should obviously be avoided. 
\end{list}
\small
\begin{verbatim}
I2:   A[] not deadlock
\end{verbatim}
\normalsize
Checking for deadlock is directly supported by Uppaal. 

\subsection{Flexibility Properties}
To guarantee that the system adapts itself dynamically to the changing traffic conditions, we define properties that allow verification of correct merges of organizations, such as the first scenario described in Fig.~\ref{fig:scenario}. In this scenario, camera 1 merges with the existing organization of camera 2 and 3. In the resulting joined organization at T1,  camera 1 is master and the other cameras are slaves. The properties for verifying correct merging are formulated as follows: 

\begin{list}{\labelitemi}{\leftmargin=1.25em}
	\item[$\bullet$] When the organization controller of a camera detects congestion (thus being master of a single member organization), then the camera merges with the neighboring organization in the direction of the traffic flow if this organization is detecting jammed traffic. Formally, we distinguish between three properties according to the current role of the neighboring camera, i.e., master with slaves, slave, and master of a single member organization. These properties are defined as follows: 
\end{list}
\small
\small
\begin{verbatim}
F1:  A<> forall(n : cam_id) 
      OrganizationController(n).CongestionDetected  
         && Camera(n+1).MasterWithSlaves  
      imply 
         Camera(n).MasterWithSlaves   
         && camera[n].slaves[n+1] 
\end{verbatim}
\normalsize

\small
\begin{verbatim}
F2:  A<> forall(n : cam_id) forall(x : cam_id) 
      OrganizationController(n).CongestionDetected  
         && Camera(n+1).Slave
         && camera[x].slaves[n+1] 
      imply 
         Camera(x).MasterWithSlaves   
         && camera[x].slaves[n] 
\end{verbatim}
\normalsize

\small
\begin{verbatim}
F3:  A<> forall(n : cam_id) 
      OrganizationController(n).CongestionDetected  
         && Camera(n+1).Master    
         && OrganizationController(n+1).CongestionDetected  
      imply 
         Camera(n).MasterWithSlaves   
         && camera[n].slaves[n+1] ||    
         Camera(n+1).MasterWithSlaves   
         && camera[n+1].slaves[n]
\end{verbatim}
\normalsize

In case the neighboring camera is master with slaves (property F1), the camera joins the organization and becomes master. In case the neighboring camera is slave (property F2), the camera joins the neighboring organization as slave. In case both are masters of single member organizations (property F3), they merge and one of them becomes master. 

\subsection{Robustness Properties}
\begin{figure}[h!]
    \centering
	 \includegraphics[width=\textwidth]{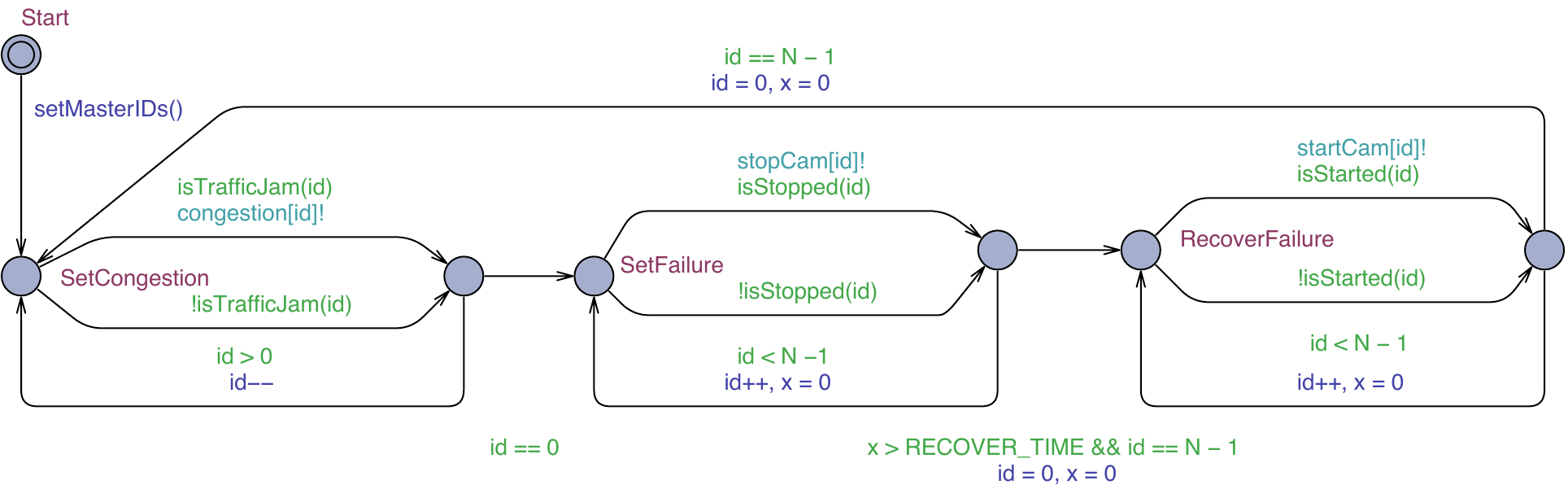}
    \caption{Environment template to inject camera failures}
    \label{fig:virtenv}
        \vspace{-8pt}
\end{figure}

To guarantee that the system recovers from a silent node failure, we define properties that allow 
verification of correct adaptations of organizations, such as the first scenario described in Fig.~\ref{fig:scenario}. In this scenario camera 2 fails at T3, which is the master of an organization with two slaves. Subsequently, the slaves elect a new master and the system recovers from the failure. 
To introduce failing cameras in the system, we modeled a virtual environment template that allows us to create sequences of traffic conditions, such as those described in Fig.~\ref{fig:scenario}. 
Fig.~\ref{fig:virtenv} shows this template. 
We focus here on properties to verify robustness of a failure of a camera in the role of master with slaves. The properties to verify robustness for failing cameras in other roles are similar. 

The scenario of Fig.~\ref{fig:scenario} is defined in the 
template's declarations section as follows: 

\small
\begin{verbatim}
   clock x;
   cam_id id = 0;

   bool isTrafficJam(cam_id id){
      if (id == 0) return false; if (id == 1) return true;
      if (id == 2) return true; if (id == 3) return true;
      if (id == 4) return false; if (id == 5) return false;
      return false;
   }

   bool isStopped(cam_id id){
      if (id == 0) return false; if (id == 1) return true;
      if (id == 2) return false; if (id == 3) return false;
      if (id == 4) return false; if (id == 5) return false;
      return false;
   }

   bool isStarted(cam_id id){
   ...

   void setMasterIDs(){
      for (i : cam_id) camera[i].m_cam = i;
      id = N-1;
   }
\end{verbatim}
\normalsize
Verifying robustness of a failing master consists of three parts:  
\begin{enumerate}
\item When the master camera fails, eventually, the self-healing controllers of the slaves detect the failure,
\item When the slaves have detected the failure, the organization controllers of the slaves will form a new organization,
\item Finally, the cameras will continue their function and monitor traffic jams. 
\end{enumerate}
To verify the correct recovering of the organization we defined the following properties:  
\begin{list}{\labelitemi}{\leftmargin=1.25em}
	\item[$\bullet$] When camera 2 fails then eventually the self-healing controllers of the slaves  detect the failure. 
\end{list}
\small
\begin{verbatim}
R1:  A<> SelfHealingController(2).Failed 
       imply 
         SelfHealingController(3).FailureDetected 
         && SelfHealingController(4).FailureDetected
\end{verbatim}
\normalsize
\begin{list}{\labelitemi}{\leftmargin=1.25em}
	\item[$\bullet$] If organization controller 2 fails and organization controller 3 or 4 has detected this (by switching to master), then eventually the organization controller of either camera 3 or 4 switches to master with slave, and the other camera becomes slave. 
\end{list}
\small
\begin{verbatim}
R2:  OrganizationController(2).Failed && 
      ((OrganizationController(3).Master 
         && OrganizationController(4).Slave) ||    
       (OrganizationController(4).Master 
         && OrganizationController(3).Slave)) 
    --> 
     OrganizationController(2).Failed && 
      ((OrganizationController(3).MasterWithSlaves 
         && camera[3].slaves[4]) ||    
       (OrganizationController(4).MasterWithSlaves 
          && camera[4].slaves[3]))
\end{verbatim}
\normalsize
\begin{list}{\labelitemi}{\leftmargin=1.25em}
	\item[$\bullet$] When camera 2 fails then eventually camera 3 and 4 will continue monitoring a traffic jam as a correct organization. 
\end{list}
\small
\begin{verbatim}
R3: A<> Camera(2).Failed 
     imply 
        ((Camera(3).MasterWithSlaves 
            && camera[3].slaves[4]) || 
         (Camera(4).MasterWithSlaves 
            && camera[4].slaves[3]))
\end{verbatim}
\normalsize
Finally, we verify whether neighbor relations\footnote{For the scenario with 6 cameras, we assigned the first camera as the neighbor of the last and vice versa.} are correctly restored after a failure: 
\small
\begin{verbatim}
R4:  A<> forall(n: cam_id) forall(x:cam_id) 
       (Camera(n).Failed && Camera(x).getLeftNeighbour() == n 
       imply 
          SelfHealingController(x).FailureDetected 
          && Camera(x).getLeftNeighbour() == n - 1)
\end{verbatim}
\normalsize
We defined a similar rule for neighbors on the right hand side.

\subsection{Design and Verification Process}

For the design of the models of the managed system (Camera, Traffic Monitor, and Organization Controller) and the managing system (Self-Healing Controller and Pulse Generator), we used Uppaal's simulator to check and correct the design. In this stage, we used the environment models (Release Traffic and Car) as shown in Fig.~\ref{fig:traffic_car} to test the different models. 
Then we defined the system invariants and the flexibility and robustness properties. To verify these properties, we designed the virtual environment template as shown in Fig.~\ref{fig:virtenv}. This restricted environment definition was needed, as verification of an arbitrary environment with randomly injected camera failures suffers from the state space problem. On the other hand, the virtual environment has the advantage that we could define the scenarios we wanted to verify with different number of cameras and increasing complexity. 

\begin{figure}[h!]
    \centering
	 \includegraphics[width=0.6\textwidth]{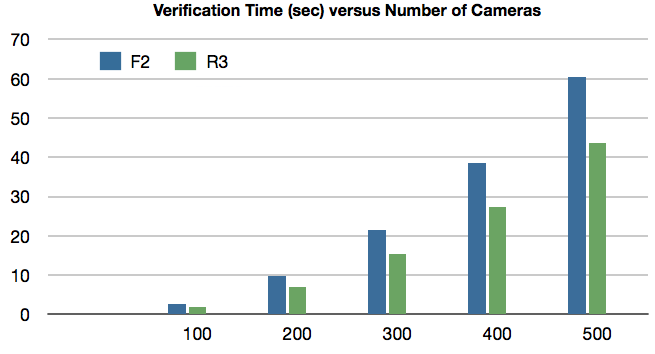}
    \caption{Verification time for increasing number of cameras}
    \label{fig:graph}
        \vspace{-8pt}
\end{figure}

To give an idea of the time required for verification, we verified the different properties for an increasing number of cameras and measured the verification time. Verification for checking that the system works properly (invariant I1) increases from 1.2 sec for 6 cameras to 197.3 sec for 60 cameras, and from 7.2 to 1330.6 sec for checking deadlock (I2). 
We also measured the verification time for flexibility and robustness properties.  Fig.~\ref{fig:graph} shows the results for properties F2 and R3. 
The figure shows that the verification time for these properties grows quasi linear with the number of cameras. 

By activating the ``diagnostic trace'' option, Uppaal can show counter examples in the simulator environment when a property is violated. This option allows to analyze the system's behavior with respect to the property and identify possible design errors. At the end of the design of the traffic monitoring system,  all properties were satisfied which was a requirement as they are defined as system requirements.  The Uppaal models of the traffic monitoring system with a prototype Java implementation  of the system is available for download via: http://homepage.lnu.se/staff/daweaa/TrafficCaseUppaal.html. 

\subsection{Discussion}

In this section, we discuss two topics. We start with explaining how the work presented in this paper fits in the integrated approach for validating quality properties of self-adaptive systems. Next, we discuss a reusable behavior model for self-adaptive systems that we derived from our study. 

As explained in the introduction, the work presented in this paper fits in an integrated approach that aims to exploit formalization results of model checking to support  model-based testing of concrete implementations and  runtime diagnosis after system deployment. To that end, it is our goal to employ the verified models to test the prototype implementation using model-based testing. 
The goal of model based testing is to show that the implementation of the system behaves compliant with this model. Model based testing uses a concise behavioral model of the system under test, and automatically generates test cases from the model. As the focus of model-based testing so far has mainly been on functional correctness of software systems~\cite{Utting2011}, and self-adaptation if primarily concerned with quality properties, we face several challenges here. A key challenges is to identify the required models to support model-based testing of quality properties.  We belief that environment models are a sine qua non for model based testing of runtime qualities, which is central to self-adaptation. In our case study, it is the environment model that specifies the failure events that have to be tested. An explicit model of the environment allows an engineer to precisely specify the failure scenarios of interest and the conditions under which the failures happens. For example, in the scenario shown in Fig.~\ref{fig:virtenv} a camera failure is generated after traffic is congested, which allows to test the correctness of the system when one of the cameras of an organization that monitors a traffic jam fails. Another challenge is to define proper test selection.
As exhaustive testing of realistic systems is typically not feasible, the tester needs to steer test selection. 
As an example, to test the self-healing scenario described in Fig.~\ref{fig:scenario}, we could mark the RecoverFailure state in the automaton shown in Fig.~\ref{fig:virtenv} as a success state. We can then formulate a reachability property to check whether the system will always reach the recover failure state after the camera has failed. This property can be issued to the model checker to test whether the implementation conforms to the model with respect to this property.

\begin{figure}[h!]
    \centering
	 \includegraphics[width=0.5\textwidth]{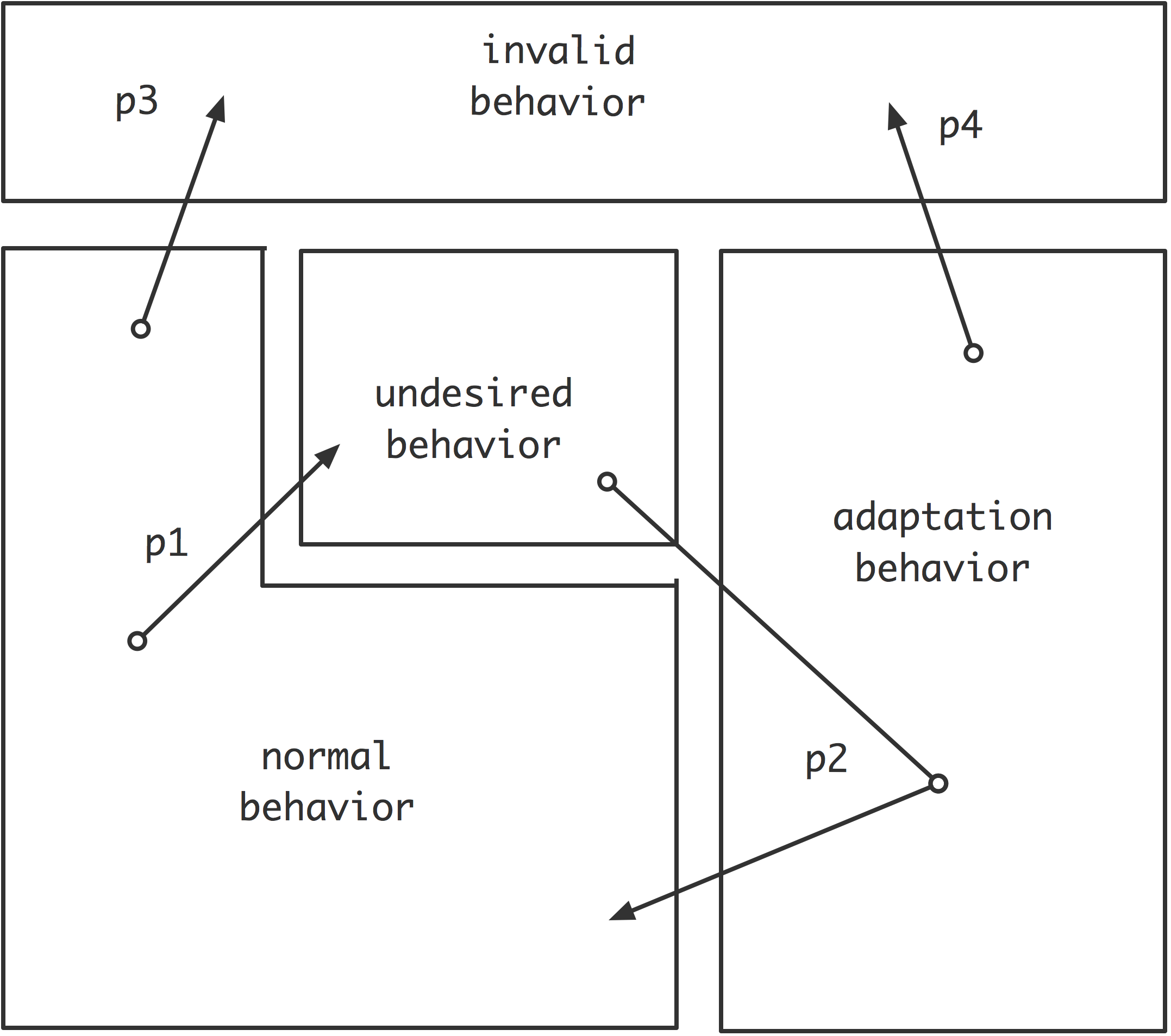}
    \caption{Zones in the state space that represent different behaviors of a self-adaptive system}
    \label{fig:zones}
        \vspace{-8pt}
\end{figure}

Finally, from our experiences with the case study, we derived an interesting model that maps the different types of behaviors of self-adaptive systems to zones of the state space. Fig.~\ref{fig:zones} shows an overview of the model. 

In the zone \textit{normal behavior}, the system is performing its domain functionality. In our case study this corresponds to monitoring traffic jams. In the zone \textit{undesired behavior}, the system is in a state where adaptation is required. In the case study, this corresponds both to a state where a reorganization is required, or a state where a camera failed. In the zone \textit{adaptive behavior}, the system is adapting itself to deal with the undesired behavior. In the case study, this means either organizations are merging or splitting, or the system heals itself from a failure. Finally, the zone \textit{invalid behavior} corresponds to states where the system should never be, e.g., deadlock in the case study. 
Properties of interest with respect to self-adaptation typically map to transitions between different zones. For example, property p1 in  Fig.~\ref{fig:zones} refers to a transition from normal behavior to undesired behavior (e.g., property R1). Property p2 refers to the required adaptation of the system to deal with the undesired behavior, that is, the system leaves the undesired state, adapts itself and eventually, returns to normal behavior (e.g., R2 to R4). Properties p3 and p4 are examples of transitions to invalid behavior that should never occur (e.g., I1 and I2).

\section{Related Work}\label{sec:relatedwork}
We discuss related work on formal modeling of self-adaptation in  three parts: fault-tolerance and self-repair, verification of various properties, and integrated approaches. We conclude with a brief discussion of the position of the work presented in this paper. 
\vspace{2pt} \\
\noindent
\textbf{Fault-tolerance and self-repair.}
\cite{Zhang2006} introduces an approach to create formal models for the behavior of adaptive programs. 
The authors combine Petri Nets modelling with LTL for property checking, including correctness of adaptations and robustness properties. \cite{Güdemann2006} presents a case study in formal modeling and verification of a robotic system with self-x properties that deal with failures and changing goals. The system is modeled as transition automata and correctness is checked using LTL and CTL (computational tree logic). 
\cite{Magee2006} outlines an approach for modeling and analyzing fault tolerance and self-adaptive mechanisms in distributed systems. The authors use a modal action logic formalism, augmented with deontic operators, to describe normal and abnormal behavior.
\cite{Ebnenasir2007} models a program as a transition system, and present an approach that ensures that, once faults occur, the fault-intolerant program is upgraded to its fault-tolerant version at run-time.  
\vspace{2pt} \\
\noindent
\textbf{Various properties.}
\cite{Becker:2006} presents a verification technique for multi-agent systems from the mechatronic domain that exploits locality. The approach is based on graph, and graph transformations, and safety properties of the system are encoded as inductive invariants. 
\cite{Khakpour2010} PobSAM is a flexible actor-based model that uses policies to control and adapt the system behavior. The authors use actor-based language Rebeca to check correctness and stability of adaptations. 
~\cite{Heinzemann2011} presents a formal verification procedure to check for correct component refinements, which preserves properties verified for the abstract protocol definition. A reachability analysis is performed using timed story charts. 
\cite{Bartels2011} considers self-adaptive systems as a subclass of reactive systems. CSP (Communicating Sequential Processes) is used for the specification, verification and implementation of self-adaptive systems. 
\vspace{2pt} \\
\noindent
\textbf{Integrated approaches.} 
\cite{Georgiadis2002} uses architectural constraints specified in Alloy as the basis for the specification, design and implementation of self-adaptive architectures for distributed systems.~\cite{Tan2006} proposes a model-based framework for developing robotic systems, with a focus on performance and failure handling. The system’s behaviour is modelled as hybrid automata, and a dedicated language is proposed to specify reconfiguration requirements. The K-Components Framework~\cite{dowling2004decentralised} offers an integrated design approach for decentralized self-adaption in which the system's architecture is represented as a graph. A configuration manager monitors events, plans the adaptations, validates them, rewrites the graph and adapts the underlying system.~\cite{Autili2009} presents the PLASTIC approach that supports context-aware adaptive services. PLASTIC uses Chameleon, a formal framework for adaptive Java applications. 
\vspace{2pt} \\
\noindent
\textbf{Position of our work.} In this paper,  we focus on modeling and verifying combined properties of flexibility and robustness of a real-world system. To that end, we use  a well-established formal method, i.e. model checking via the Uppaal tool. Most existing formal approaches for self-adaptive systems assume a central point of control to realize adaptations. We target systems in which control of adaptations is \textit{decentralized}, that is, managing systems detect the need for adaptations and coordinate to realize the required adaptations locally. Most researchers employ formal methods in one stage of the software life cycle; notable exceptions are~\cite{Georgiadis2002,Zhang2006,Tan2006}. The formal approach used in this paper supports architectural  design, but fits in an integrated formally founded approach to validate the qualities of self-adaptive systems that aims to exploit formal work products during subsequent stages of the software life cycle.

\section{Conclusions and Challenges Ahead}\label{sec:conclusions}

In this paper, we presented a case study on formal modeling and verification of a decentralized self-adaptive system. The Uppaal tool allowed us to model the system and verify the required flexibility and robustness properties. We defined a dedicated environment model both to verify specific adaptation scenarios and manage the state space problem.
This work fits in our long term research objective to develop an integrated approach for formal analysis of decentralized self-adaptive systems that combines verification of architectural models with model-based testing of applications to guarantee the required runtime qualities. 
As the next step in our research, we plan to build upon the work presented in this paper in two ways. First, we plan to study how we can apply verified architecture models to test concrete implementations using model based testing~\cite{Utt:Leg07}. As model based testing has mainly focused on functional testing so far, a key challenge here is to extend the approach to test quality attributes. 
Second, we plan to elaborate on the initial zone-based model of self-adaptive systems that defines the different types of behavior of this class of systems. To that end, we are currently performing a systematic literature review on model checking of self-adaptive systems to map existing work on the model.

\bibliographystyle{eptcs}
\bibliography{FMSAS2012.bib}
\end{document}